\documentclass{article}

\usepackage{PRIMEarxiv}

\usepackage[utf8]{inputenc} % allow utf-8 input
\usepackage[T1]{fontenc}    % use 8-bit T1 fonts
\usepackage[colorlinks,allcolors=blue]{hyperref}       % hyperlinks
\usepackage{url}            % simple URL typesetting
\usepackage{booktabs}       % professional-quality tables
\usepackage{amsfonts}       % blackboard math symbols
\usepackage{nicefrac}       % compact symbols for 1/2, etc.
\usepackage{microtype}      % microtypography
\usepackage{lipsum}
\usepackage{fancyhdr}       % header
\usepackage{graphicx}       % graphics
\graphicspath{{media/}}     % organize your images and other figures under media/ folder

\title{Code Code Evolution: Understanding How People Change Data Science Notebooks Over Time}

\author{
  Deepthi Raghunandan \\
  University of Maryland \\
  College Park \\
  Maryland \\
  USA \\
  \textit{draghun1@umd.edu}
   \And
  Aayushi Roy \\
  University of Maryland \\
  College Park \\
  Maryland \\
  USA \\
  \textit{aroy2530@umd.edu}
   \And
  Shenzhi Shi \\
  University of Maryland \\
  College Park \\
  Maryland \\
  USA \\
  \textit{sshi1234@umd.edu}
     \And
  Niklas Elmqvist \\
  University of Maryland \\
  College Park \\
  Maryland \\
  USA \\
  \textit{elm@umd.edu}
     \And
  Leilani Battle \\
  University of Washington \\
  Seattle \\
  Washington \\
  USA \\
  \textit{leibatt@cs.washington.edu}
}

\begin{document}

\maketitle

\begin{abstract}
    \emph{Sensemaking} is the iterative process of identifying, extracting, and explaining insights from data, where each iteration is referred to as the ``\emph{sensemaking loop}.''
    Although recent work observes snapshots of the sensemaking loop within computational notebooks, none measure \emph{shifts} in sensemaking behaviors over time---between exploration and explanation.
    This gap limits our ability to understand the full scope of the sensemaking process and thus our ability to design tools to fully support sensemaking.
    We contribute the first quantitative method to characterize how sensemaking evolves within data science computational notebooks.
    To this end, we conducted a quantitative study of 2,574 Jupyter notebooks mined from GitHub.
    First, we identify data science-focused notebooks that have undergone significant iterations.
    Second, we present regression models that automatically characterize sensemaking activity within individual notebooks by assigning them a score representing their position within the sensemaking \textit{spectrum}. 
    Finally, we use our regression models to calculate and analyze shifts in notebook scores  across GitHub versions.
    Our results show that notebook authors participate in a diverse range of sensemaking tasks over time, such as annotation, branching analysis, and documentation.
    Finally, we propose design recommendations for extending notebook environments to support the sensemaking behaviors we observed.
\end{abstract}

% keywords can be removed

\section{Introduction}
\label{sec:intro}

Sensemaking is ``the process of searching for a representation and encoding data in that representation to answer task-specific questions''~\cite{Russell1993}. 
In each iteration of this ``sensemaking loop''~\cite{pirolli2005sensemaking}, data scientists refine their code, visualizations, and annotations in pursuit of a deeper understanding of their data~\cite{Thomas2005}.
In the process, data scientists often oscillate between \emph{exploring} the data and \emph{explaining} what they have learned (to themselves or stakeholders)~\cite{rule2018exploration,pirolli2005sensemaking}, leading to more of ``spiral'' of activity than a true ``loop''.

Computational notebooks such as Jupyter~\cite{kluyver2016jupyter}, R Markdown, or Observable are especially popular for documenting the complexities of the sensemaking process given the ease with which code can be interleaved with descriptive text and illustrative images~\cite{kery2018story, rule2018exploration}.
However, notebooks still fall short of the ideal for sensemaking, particularly in tracking changes to notebooks over time~\cite{kery2017variolite}, frustrating many notebook users~\cite{chattopadhyays2020pain}.

In order to improve notebooks for sensemaking we must first characterize users' common interaction patterns so that we can (re)design notebook environments to better support them~\cite{kery2017variolite}.
However, the evolving nature of sensemaking suggests that these patterns may vary depending on where users are within the progression between \emph{exploration} and \emph{explanation}~\cite{rule2018exploration}.
Thus, we need to determine where a user is along this exploration-explanation spectrum before we can design appropriate solutions. 
Recent work posits that we can infer where a user is within the exploration-explanation spectrum directly from computational notebooks~\cite{kery2018story, rule2018exploration, wang2020assessing}.
However, these prior works rely on small-scale user studies to investigate sensemaking within notebooks. 
Furthermore, they treat notebooks as static outputs of sensemaking rather than a core medium for iteration.
This limits our understanding of notebooks as living documents of scientific inquiry.
Without more rigorous validation, it is still unclear whether current theory can accurately detect sensemaking within real-world notebook environments.

This paper proposes a new approach to analyzing how computational notebooks are revised over time.
The key idea is that many analysts already track their notebook iterations using public version control infrastructure such as GitHub.
We contribute a pipeline to collect, model, and quantify sensemaking behaviors across GitHub commits.
This pipeline allows us to (1) characterize observed shifts in sensemaking behavior within notebooks, such as whether notebooks become more explanatory or exploratory over time, and to (2) understand why these shifts occur.
To do this, we randomly sampled and downloaded 2,574 Jupyter notebooks stored on GitHub's public repositories.
We report on their content, revision history, and evolution. Our analysis has three parts:

\begin{enumerate}

    \item\textbf{Identifying Relevant Notebooks} -- finding the data science notebooks that were actively refined overtime on GitHub, as well as quantitative metrics to analyze them;
    
    \item\textbf{Measuring Exploration vs.\ Explanation} -- leveraging prior work~\cite{rule2018exploration, kery2018story} to distinguish between the exploratory vs.\ explanatory nature of data science notebooks; and
    
    \item\textbf{Measuring Evolution} -- drawing on GitHub revision history to understand how notebooks, and in turn authors' positions in the sensemaking loop, shifted over time.
    
\end{enumerate}

We acknowledge that these quantitative approximations may not reflect the author's complete sensemaking process.
This discrepancy is due in part to authors' selective reporting as well as to limitations inherent to notebook platforms themselves~\cite{liu2020paths, wang2019collab, muller2019human, kery2018story, wang2020assessing, rule2018exploration, chattopadhyays2020pain}.
In other words, the use of GitHub as a data source likely biases the type of notebooks we collect for our sample.
Regardless, we believe that measuring notebooks as they publicly change over time still provides a unique perspective on the sensemaking process that qualitative analyses of singular notebook versions cannot.

We make the following contributions in this paper:
(1) we develop a rubric to show how to quantify the explanatory or exploratory nature of a Jupyter Notebook, enabling us to analyze data-science notebooks at scale;
(2) we track the evolution of notebooks over time by calculating our quantitative measure across multiple notebook versions;
(3) we characterize the way analysts iterate on their notebooks; and
(4) we use these insights to make design recommendations to better support the different notebook-based sensemaking behavior we observed.
More broadly, we contribute a more nuanced view of the data science process that brings notebook analysis methodology closer alignment with established theories of sensemaking and data exploration.
This, in combination with a quantitative approach, can be directly applied towards teaching, guiding, and implementing tools for best practices in data science.

Beyond the overview of our method and results presented in this paper, we have also provided supplementary material with the full details in the following anonymous OSF repository: \url{https://osf.io/9q4wp/?view_only=61e6f58d29194742a0aaed328afdea4d}

%% ---------------------------------------------------------------------
%% Related Work
%% ---------------------------------------------------------------------
\section{Related Work}
\label{sec:related-work}

In this section, we introduce key concepts and terminology that we use in our work to map signs of exploration and explanation in computational notebooks to corresponding shifts within the sensemaking loop.

% We will review key terminology and concepts from the literature that inform our analysis structure. First, we examine our current understanding of data science. Then we outline how this understanding influences how data scientists use computational notebooks. This work uniquely builds on this understanding by looking versioned Jupyter Notebooks documents.

% Prior work has found that data science involves discovery, cleaning, profiling, analyzing, and communicating data~\cite{KandelPHH12, Rule2018PhD, guo2012software}. 
% Data science is characterized by sensemaking, which is the process of iterating on and between tasks to better understand and represent the data~\cite{Russell1993}. 

\subsection{Computational Notebooks}

As an embodiment of the literate programming paradigm~\cite{knuth1984literate}, where traditional source code is embedded in descriptive natural language, \textit{computational notebooks} are an ideal medium for studying the sensemaking process~\cite{colaboratory, observable, kluyver2016jupyter, radle2017codestrates, tabard2008individual, Wickham2017}.
A notebook is a linear sequence of executable code that perfectly captures the procedural nature of sensemaking.
The ability to inspect intermediate results by generating visualizations and tables scaffolds the exploratory process.
The rich annotation features scaffold the pivoting of data representations towards explanation.
Notebooks also allow for easy sharing of data, code, and analyses all in one~\cite{rule2018exploration, chattopadhyays2020pain}.
As a result, computational notebooks have quickly become an essential part of conducting data science~\cite{kery2018story, kluyver2016jupyter, rule2018exploration, wang2019collab}.

Data scientists utilize computational notebooks---specifically their flexible cell structure---to iterate on different branches of exploration and create narratives surrounding their analyses~\cite{kery2018story, rule2018exploration, wang2021makes, wang2022documentation, chattopadhyays2020pain, pimentel2019large, wenskovitch2019albireo, koenzen2020code, kallen2020jupyter, dong2021splitting, rehman2019towards, wang2020assessing}. 

\subsection{Sensemaking in Computational Notebooks}

Pirolli and Card describe the sensemaking loop as cumulative iterations by which analysts develop an understanding of the data~\cite{pirolli2005sensemaking}.
Each iteration informs the next.
Our work enriches this definition of sensemaking with notebook-oriented notions of exploration and explanation from the literature~\cite{rule2018exploration, kery2018story}.
Specifically, we define the ``sensemaking spectrum'' as a two-dimensional representation of the sensemaking loop from early-stage exploration to late-stage explanation~\cite{pirolli2005sensemaking}.
% We refer to the sensemaking ``spectrum'' and ``loop'' interchangeably. 

Qualitative studies from Rule et al.\cite{rule2018exploration}, Kery et al.~\cite{kery2018story}, and Wang et al.~\cite{wang2020assessing} show that we can observe sensemaking within computational notebooks in the form of exploration and explanation. 
For example, Kery et al.\ observed that some analysts created many small code cells while performing exploratory data analysis to optimize iteration, and later grouped code into individual ``logical units'' to communicate analytical steps~\cite{kery2018story}. 
Rule et al.\ noted that analysts place text that serve different purposes in different parts of the notebook~\cite{rule2018exploration}.
They found that nearly all code comments help explain the methods employed by code, headers labeled the analyses, and most non-header text explained analytical steps. 
Wang et al.\ extended this finding by showing that highly readable (explanatory) notebooks use a variety of descriptors to attract a broader audience~\cite{wang2021makes}.
Based on these findings, it seems evident that different notebook characteristics, such as types of documentation or distribution of code across cells, can indicate an analyst's current position within the sensemaking spectrum between exploration and explanation. 

Some previous work contribute to our understanding of how notebooks are used but do not identify these steps in the context of sensemaking.
For example, Dong et al.\ find that code cleaning is an integral part of sharing a notebook~\cite{dong2021splitting}.
They characterize cleaning as renaming variables, generating functions, reordering code cells, adding pertinent annotations, moving content between files, and removing extraneous content.
We leverage Dong et al.'s work to construct a comprehensive model of sensemaking in notebooks. 

\subsection{From Exploration to Explanation}

In data \emph{exploration}, analysts seek to profile their data, define their goals, and become comfortable with potential analytic methods~\cite{kery2018story, pimentel2019large, guo2012software}. 
As Alspaugh et al.\ explain, data analysis exists within a spectrum between ``exploratory" and ``directed" analysis, wherein the nature of analysis changes as goals become more concrete~\cite{alspaugh2018futzing}. 
Analysts seek to understand their dataset, look for exciting patterns, and identify assumptions as a means to inform next steps~\cite{kandel-enterprise, alspaugh2018futzing, muller2019data, wongsuphasawat-eda}.
The ultimate objective of this process is to inform decisions. 

%\subsection{Data Explanation}

The process of \emph{explaining} data insights entails shaping explorations into a narrative to communicate the process and results~\cite{kery2018story}.
This type of explanation provides clarity on how the analysis process yielded particular insights to an audience (including oneself).
Analysts can describe their analyses and findings in varying levels of detail and clarity, ranging from reporting all avenues of exploration and ensuing insights to saving only the most critical decisions and findings~\cite{kery2018story, MathisenHKGE19}.
The level of detail they choose depends mainly on the audience.
When the audience is oneself or fellow technical team members, the analyst focuses on retaining code and branches of exploration and formatting them in a comprehensible manner~\cite{rule2018exploration, kery2018story}.
When presenting results to a broader, perhaps non-technical, audience, analysts may remove details that appear confusing or uninteresting and add more explanatory text---shifting the focus from the code to the narrative~\cite{liu2020paths, rule2018exploration, kery2018story}.

Our definitions are also grounded in literature on notebook reproducibility---a common motivation for authoring notebooks~\cite{pimentel2019large, wang2020assessing, rehman2019towards, wenskovitch2019albireo, chattopadhyays2020pain}.
Like explanatory notebooks, reproducible notebooks allow for communication, reuse and reproduction---enabling a clear linear structure~\cite{rehman2019towards, wenskovitch2019albireo} and presenting clean code~\cite{dong2021splitting, pimentel2019large}.
%We incorporate these findings to a rubric of definitions pertaining to notebooks in the sensemaking spectrum. We use existing definitions to quantify a user's position within the sensemaking loop

%% ---------------------------------------------------------------------
%% Data Collection and Filtering
%% ---------------------------------------------------------------------
\section{Dataset}
\label{sec:dataset}

Given our aim to use revision histories to measure how data science notebooks evolve, we chose to analyze publicly available Jupyter Notebooks found on GitHub for the following reasons: 1) it provides an extensive repository of Jupyter Notebook documents, and 2) they are particularly amenable to meta-analysis due to the ease of accessing their underlying JSON metadata structure.

In July 2019, we identified 4.7 million notebooks on GitHub and downloaded a random sample of 60,000. 
Using the above criteria, we selected 2,574 notebooks of these for further analysis.

We queried for \textit{GitHub commit} information to ensure we could examine all versions of notebooks.
We anticipated that, given that GitHub contains a large variety of notebooks in terms of quality and purpose, not all notebooks would be suited for this analysis.
Thus, we sought to programmatically filter for data science notebooks to automatically scale our analysis to any sample size.
Here, we describe our methods for collecting, sampling, and filtering notebooks for subsequent meta-analysis.

\subsection{Data Collection Method}
\label{sec:approach:selection}

To mine Jupyter Notebooks from GitHub, we used Rule et al.'s approach~\cite{rule2018exploration}.
We first downloaded and accessed a total of 59,887 notebooks. 
For the sake of project feasibility, we chose to observe only Python notebooks annotated in English.
Python and R are the most popular languages used for data science, but we made this choice on the basis that Python is significantly more common in Jupyter than R (more than 96\% of all Jupyter Notebooks are written in Python~\cite{rule2018exploration}).
This eliminated 3,642 notebooks from our sample.

To further select notebooks suitable for our analysis, we defined a standard to identify Jupyter notebooks that use \textbf{data science} and were \textbf{stored on GitHub}.
To meet our standards: 1) notebooks must demonstrate data analysis activity, 2) some subset of changes must be observable across multiple versions found in the GitHub repository, and 3) changes to the notebook must be made by the original owner of the notebook.
We briefly outline our filtering criteria below; further details about our criteria and our methods can be found in our supplemental material.

\paragraph{Data Science Notebooks}

First, we used the number of popular data science Python libraries as a heuristic to determine whether the notebooks were data science-oriented.
We manually derived a list of data science libraries by reviewing 28 online Python tutorials.
The full list contains the following packages:
\texttt{numpy}, \texttt{scipy}, \texttt{pandas}, \texttt{scikit-learn}, \texttt{matplotlib},  \texttt{pytorch}, and \texttt{tensorflow}~\cite{pimentel2019large}.
We searched and filtered for notebooks with these libraries and the associated API calls. 
A total of 35,692 notebooks out of 59,887 were eliminated for not meeting this criterion.

\paragraph{Versioned Notebooks}

Second, we used the number of notebook versions within the repositories and the number of changes within them as a heuristic to measure whether changes to data analysis were observable.
We found that we could observe changes to notebook cells and lines when there were at least 4 revisions.
We required that notebook versions contained at least 2 additions or deletions of cells, along with at least 20 additions or deletions to lines across cells.
A total of 5,082 were eliminated for not meeting this criterion.

\paragraph{Original Content}

When there are multiple contributors to a notebook, it is unclear whether both authors had the same intent in making changes.
We restrict our analysis to only include notebooks updated by the owner of the notebooks' respository to eliminate this ambiguity.
An additional 12,350 notebooks were eliminated based on this criterion.

\subsection{Summary \& Dataset Considerations}

2,574 notebooks passed the above criteria.
A majority of these notebooks were data-science oriented, contained a rich amount of data science activity, and were written in English.
We acknowledge that our quantitative approach may not be perfectly accurate.
Even with our rigorous criteria, the dataset may still contain notebooks that do not demonstrate traditional sensemaking.
We believe these will be in the minority given our threshold on data science API usage combined with the precision of our commit filters.

We also acknowledge that not all analysts may commit all their iterations to GitHub repositories.
However, given the relatively large size of our dataset (2,574 notebooks and 26,474 notebook versions), the noise in our dataset is significantly diminished.

%% ---------------------------------------------------------------------
%% Study 2
%% ---------------------------------------------------------------------
\section{Measuring Exploration vs.\ Explanation}
\label{sec:study2}

We first seek to answer the following research question:
\textit{Can we apply previous findings to quantitatively measure exploration and explanation in computational notebooks?}
To answer this question, we first manually curated a reference dataset of 244 notebooks (10\% of our sample) using a manual rubric that maps notebook characteristics to points within the sensemaking spectrum.
This rubric scores builds directly upon findings of previous work regarding how sensemaking manifests within computational notebooks. 
We then used our manual reference dataset to develop and validate a model to automate the manual classification. 

\subsection{Constructing the Reference Dataset}
\label{sec:study2:ground-truth}

Here, we describe a new rubric to score data science notebooks according to their position along the sensemaking spectrum, which we aim to automate.
The goal of this analysis was to determine whether each notebook appears to be more exploratory or explanatory in nature.
As observed in prior work~\cite{rule2018exploration, kery2018story, wang2021makes}, notebooks with good narrative structure generally tell a compelling story of both the data analysis process and the insights derived from this process.
These notebooks clearly communicate the analyst's motivations and insights, and can appeal to a wide audience via instructional text or explanations of field-specific terminology~\cite{alspaugh2018futzing, chattopadhyays2020pain}.
Therefore, the better the narrative structure of a notebook, the higher the sensemaking score it should receive.

Based on the literature, one of the authors developed a rubric for scoring notebooks.
Two other authors provided feedback on the rubric between iterations.
All coders were knowledgeable in data science.
Inspired by methods for reaching agreement on qualitative codes~\cite{mcdonald2019reliability}, our scoring process involved three iterations with one coder to converge to scores consistent with our rubric.
The iterations included a preliminary scoring iteration where scores were assigned based on an initial rubric, a second one where scores were refined in parallel with the rubric, and a final one where scores were reviewed for consistency with the final rubric.

We represented the positions of notebooks on the sensemaking spectrum using a score between 0.1 and 1.0, where 0.1 represents the most exploratory notebooks and 1.0 the most explanatory ones.
Scores were assigned in increments of 0.1.
We wanted our range of scores to be evenly distributed such that the first five scores (0.1-0.5) characterize mostly exploratory notebooks and the last five scores (0.6-1.0) characterize mostly explanatory notebooks.

To form an impression of the notebooks, coders considered the following criteria: 
code abstraction methods such as functions, classes, and code distribution within cells; the clarity of code based variable names and in-line code comments; the use of markdown headers to create sections; the cohesiveness of the analytical workflow; and the types of descriptors (analytical, procedural and context) included in the document.
We identify exploratory notebooks as ones which leverage code to explore data.
These notebooks place little focus on explaining insights, reasoning, or analytical methods, and instead focus on fast iteration resulting in duplicated, messy code~\cite{kery2018story, chattopadhyays2020pain, koenzen2020code}. 
In explanatory notebooks, on the other hand, the intent to outline, document, or explain previous data exploration is clear~\cite{rule2018exploration, wang2021makes, chattopadhyays2020pain, alspaugh2018futzing}.
The rubric in \autoref{tab-rubric} specifies how each notebook was evaluated. 
Coders reached a final inter-rater reliability score of $0.88$ after converging on this rubric.

\subsection{Automating the Scoring Process}

To scale up our analysis, we needed a way to programmatically calculate a notebook's sensemaking score.
We observed characteristics of notebooks from our reference set to understand how they contributed to a notebook's position on the sensemaking spectrum. 
We chose to observe particularly quantitative characteristics which were highlighted by previous literature on sensemaking in notebooks and used our own observations to understand how other metrics correlated with our manually assigned sensemaking scores.

\subsubsection{Analyzing Notebook Characteristics}

We analyzed all parts of a notebook as follows.

\paragraph{Code}

Given that notebook authors perform their analyses primarily through code, we were interested in how an author's approach to writing their code may contribute to the exploratory nature of the notebook. 
The amount of code in the notebook tends to increase as notebook authors explore their dataset~\cite{head2019messes, kery2018story, rule2018exploration, pimentel2019large, chattopadhyays2020pain, dong2021splitting, kallen2020jupyter, koenzen2020code}. 
To this end, we computed the following measures for each notebook in our reference set: the number of code cells and the total lines of code across all code cells.

\paragraph{Non-Code}

Data science notebooks often contain content beyond just code, such as markdown text, table outputs, text outputs, visualization outputs, and code comments.
This content is of particular interest, because they are primarily explanation-oriented, and thus increase a notebook's sensemaking score~\cite{rule2018exploration, wang2022documentation, rule2018ten, kery2018story}.
To gauge the impact of non-code content within notebook cells, we computed the following measures: the number of markdown cells; the total lines across markdown cells; the number of text outputs produced by code cells; the number of tables produced by code cells; the number of visualizations produced by code cells; and the number of individual code comments within code cells.

\paragraph{Whitespace}

Negative space can have a profound impact on how information is organized and presented for communication~\cite{tufte2001visual}.
Notebook authors can control the negative space in their notebooks by adding newline and space characters.
Rule et al.\ suggest that the number of spacing characters in text and code cells could point towards more of an explanation focus for a notebook~\cite{rule2018exploration}.
Therefore, we also measure the total spacing characters observed across code cells, and spacing characters across markdown cells.

\subsubsection{Measure Normalization}

To ensure that different quantitative measures could be compared fairly across notebooks, we normalized each measure with respect to individual notebooks.
Our normalization process translates each measure to a domain of 0.0 to 1.0.
We normalized cell counts by dividing them by the total number of cells found in the notebook (e.g., the number of code cells were divided by the number of all cells).
We normalized the output (text, table, and visualization) counts by dividing them by the total number of outputs in the notebook.
We normalized the number of code comments by the number of lines found within the code cell.
Finally, we normalized the number of spaces for a given cell type by dividing by the total spaces across all notebook cells.

\subsubsection{Combining Measures}

We tested two different combinations of the measures based on our understanding of how notebooks are used in practice.
The first combination focused on the outputs generated by a notebook, which may indicate a more explanatory notebook.
This combination consists of total markdown cells, total tabular outputs, total visualization outputs, and total text outputs.
We refer to this combination as ``Output-Focused.''

Rather than emphasizing the outputs of code cells, our second combination of measures gauged the proportions of different cell types and their structure, where notebooks with more markdown cells and/or better organized cells were likely to be more explanatory.
This combination includes the following measures: total code cells; total markdown cells; total spaces in markdown cells; and total lines in markdown cells.
We refer to this combination as ``Organization-Focused.''

\paragraph{Synthesizing a ``Hybrid''}

It is possible that cell outputs and cell organization together play important roles in assessing the sensemaking score of a notebook.
In response, we formulated a new ``Hybrid'' combination, incorporating measures from both of the above combinations: total code cells; total markdown cells;  total markdown spacing; total text outputs; total visualization outputs; and total tabular outputs.
We refer to this combination as ``Hybrid.''

\subsubsection{Comparing Combinations of Quantitative Measures}

We used each combination of measures as parameters in a multi-linear regression analysis against the manually assigned sensemaking scores.
We leveraged a $k$-fold cross-validation technique to ensure the strength of each model.
The models were trained in 5 folds on 20\% of the data, and tested on the rest.
$R^2$ values were calculated for each model, within each fold.
A mean and median $R^2$ value were generated for each model.
Median $R^2$ values were compared to assess correlations.

\subsubsection{Hybrid-Focused Combination Performance}
\label{sec:study2-hybrid}

We found that our Output-Focused and Organization-Focused combinations correlate positively with increases in sensemaking score.
The Organization-Focused combination has a higher correlation value than our Output-Focused combination. 

However, it is unclear whether cell outputs and cell organization measure redundant information, or are complementary.
To assess this relationship, we performed the same analysis with our ``Hybrid'' combination, which produces a multi-linear regression model with a correlation value ($R^2=0.591$):
$Y = 0.426 \times \mathtt{totalMarkdownCells} + 0.145 \times \mathtt{totalMarkdownSpace} - 0.077 \times \mathtt{totalCodeCells} + 0.176 \times \mathtt{totalVisualizations} + 0.125 \times \mathtt{totalTextOutputs} + 0.172 \times \mathtt{totalTableOutputs} + 0.395$.

\begin{figure}
    \centering
    \includegraphics[width=0.5\columnwidth]{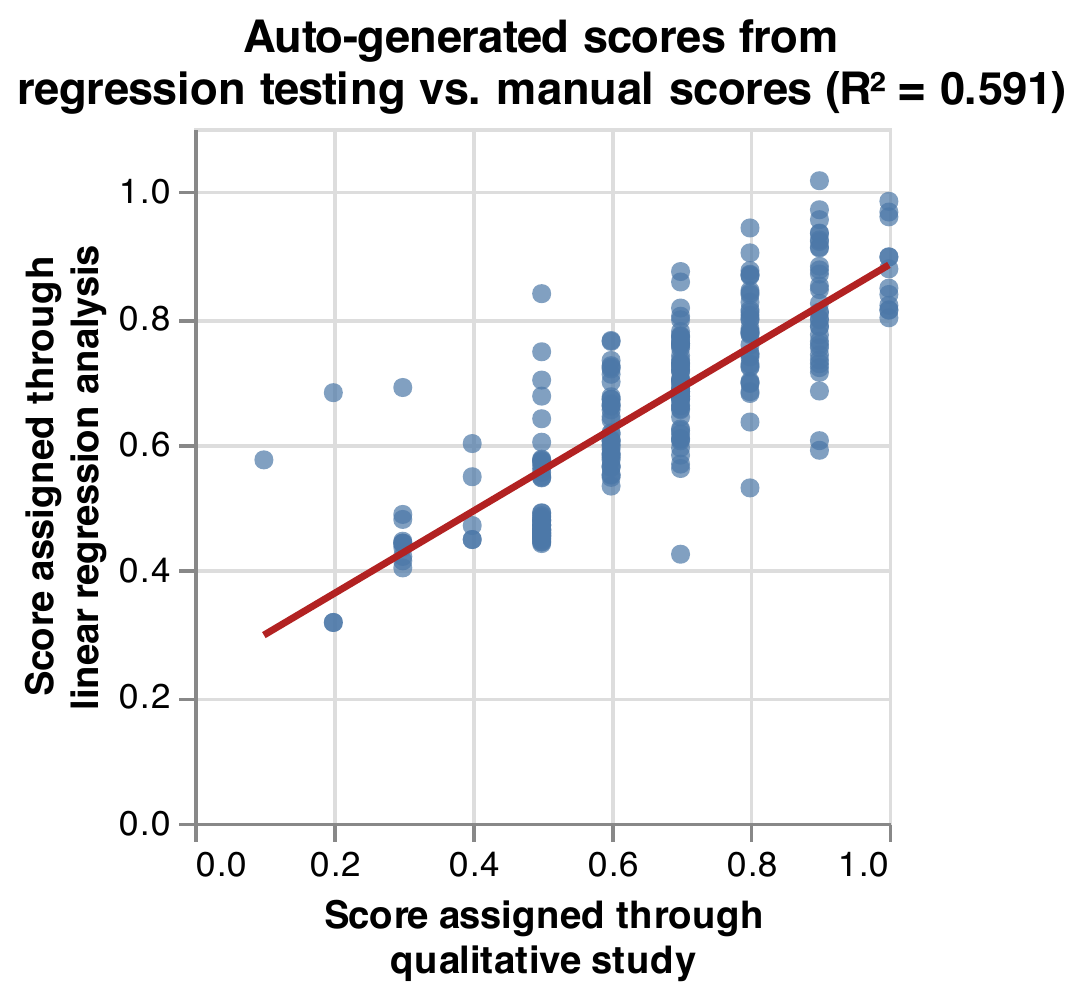}
    \caption{\textbf{Score comparison.}
    Comparison of our manually assigned sensemaking scores (x-axis) and hybrid automated scores (y-axis), using a combination of quantitative measures.
    These data points are from our reference-set containing 244 notebooks.
    (Multi-linear regression model, $R^2=0.591$.)} 
    \label{fig:study2-manual-versus-automated}
\end{figure}

Thus, it seems that cell types, cell outputs, and cell organization capture separate but complementary facets of a notebook author's sensemaking process.
For this reason, \textbf{we use the hybrid combination for all subsequent analyses} in this paper.
The results for the ``Hybrid'' combination are provided in \autoref{fig:study2-manual-versus-automated}, where the y-axis represents the range of automated sensemaking scores, and the x-axis the manually assigned sensemaking scores.

\subsection{Results and Takeaways}

In this study, we used prior observations of how analysts interact with the sensemaking process as a whole~\cite{pirolli2005sensemaking, kery2018story, kross2019practitioners, Russell1993, alspaugh2018futzing, kandel-enterprise} and computational notebooks in particular~\cite{rule2018exploration} to derive a new rubric for assigning a sensemaking score to an individual notebook.
To scale up the application of our rubric, we first formed a manually-labeled reference dataset containing 244 notebooks.
These were used to develop our regression models.
We then analyzed relevant quantitative measures that may predict these sensemaking scores.
We found a strong correlation between increases in sensemaking scores and increases in organization-focused measures, such as having more lines or more spacing in markdown cells.
We also found a positive correlation between output-focused measures and sensemaking score (i.e., more explanatory notebooks).
Thus, the content and structure of a notebook may be indicative of the sensemaking goals of the notebook's author. 

We acknowledge that it is unrealistic to extract exact quantitative measures for sensemaking scores and that many combinations of notebook attributes could ultimately predict sensemaking behavior. 
However, our scoring mechanism aims to establish sufficient placement along the exploration-explanation spectrum, allowing us to observe shifts in the spectrum over time. 
Although not exact, our computed scores still provide a valuable signal for studying notebook evolution.
We encourage the community to extend our initial feature set with new attributes in Section~\ref{sec:discussion}.

%\changed{During the rubric development process, coders found that each sensemaking scores could not be perfectly mutually exclusive from adjacent scores given that different authors leverage different methodologies to encode a workflow narrative. Given that our ultimate aim is to observe sensemaking evolve, we do not believe that this type of rubric introduces major limitations. 

%\changed{Coders also found that the notebook attributes they observed were only a subset that could capture sensemaking. In observing an R value above 0.5, we have shown that even a small subset of notebook characteristics can map to sensemaking behavior. Exploring other attributes and creating more definite categories within the rubric would be an important part of future work.}

\begin{table*}[htbp]
    \centering
    \begin{tabular}{cp{5cm}p{10cm}}
    % \begin{tabular}{cp{4cm}p{9cm}}
    \toprule
    \textbf{Score} & \textbf{Stage} & \textbf{Concrete Examples} \\
    \midrule
    0.1 &
    Understanding the data~\cite{kandel-enterprise, muller2019data, pirolli2005sensemaking, wongsuphasawat-eda, alspaugh2018futzing} &
    Just code (Few unorganized cells~\cite{kery2018story, rule2018exploration}, duplicated code~\cite{kallen2020jupyter}, no output~\cite{pimentel2019large}.)
    \\
    0.2 &
    Iterative data wrangling~\cite{kandel-enterprise, guo2012software, muller2019data, pirolli2005sensemaking, wongsuphasawat-eda, alspaugh2018futzing} &
    Just code, some output from exploration~\cite{pimentel2019large} (disjoint, duplicated code cells; lots of code cells with individual lines of code~\cite{kery2018story, rule2018exploration, head2019messes, chattopadhyays2020pain}).
    \\
    0.3 &
    Defining goals using iterative exploratory analysis~\cite{kandel-enterprise, alspaugh2018futzing, wongsuphasawat-eda, pirolli2005sensemaking} &
    Lots of code~\cite{kery2018story}, some output from exploration (some code cells are grouped by functionality, text headers and code comments are used to label groups)~\cite{kery2018story, rule2018exploration, dong2021splitting, head2019messes, chattopadhyays2020pain}.
    \\
    0.4 &
    Beginning goal-oriented exploratory analysis~\cite{kandel-enterprise, alspaugh2018futzing, wongsuphasawat-eda, pirolli2005sensemaking} &
    Code and visual output address some goals (some code cells are linearly grouped by text headers and code comments are used to label and annotate analysis)~\cite{rehman2019towards, wenskovitch2019albireo, kery2018story, head2019messes, dong2021splitting, rule2018exploration, wang2020assessing}.
    \\
    0.5 &
    Exploratory analysis with clear goals~\cite{kandel-enterprise, alspaugh2018futzing, wongsuphasawat-eda, pirolli2005sensemaking} &
    Code and visual output address goals (majority of code cells are linearly grouped by text headers and code comments)~\cite{rehman2019towards, wenskovitch2019albireo, kery2018story, head2019messes, dong2021splitting, rule2018exploration, wang2020assessing}.
    \\
    0.6 &
    Analytical steps are communicated~\cite{kandel-enterprise, alspaugh2018futzing, wongsuphasawat-eda, pirolli2005sensemaking, guo2012software} &
    Code and code output are interwoven with text headers and code comments for the sake of outlining the logical steps which were taken.~\cite{kery2018story, head2019messes, rule2018exploration, wang2021makes}.
    \\
    0.7 &
    Some insights of analysis tracked and communicated~\cite{kandel-enterprise, alspaugh2018futzing, wongsuphasawat-eda, pirolli2005sensemaking, guo2012software} &
    Analysts are also using text to briefly annotate their code with insights from individual logical steps.~\cite{wang2021makes, rule2018exploration, kery2018story}.
    \\
    0.8 &
    Some motivations and insights tracked and communicated~\cite{kandel-enterprise, alspaugh2018futzing, wongsuphasawat-eda, pirolli2005sensemaking, guo2012software} &
    Analysts are explaining their motivations behind individual analytical steps, and more thoroughly annotating their logical steps with insights. ~\cite{rule2018exploration, wang2021makes, kery2018story}.
    \\
    0.9 &
    Motivations and insights of analysis clearly communicated~\cite{kandel-enterprise, alspaugh2018futzing, wongsuphasawat-eda, pirolli2005sensemaking, guo2012software} &
    Analysts introduce their analytical reasoning behind the work overall.
    In addition, they are illustrating links between logical steps using text in the form of headers, insights and motivations.
    Together the text forms a narrative of the methods and the results.~\cite{rule2018exploration, wang2021makes, kery2018story}.
    \\
    1.0 &
    Analysis workflow communicated to a wide audience~\cite{pirolli2005sensemaking} &
    Text may provide instructions on how to interact with the notebook, provide context behind the work, motivations on the methodology, insights from individual logical steps and insights from the entire exercise~\cite{rule2018ten}. If code and code output are present, they align with the narrative being outlined by the text.~\cite{wang2021makes, rule2018exploration, kery2018story}.
    \\
    \bottomrule
    \end{tabular}
    
    \caption{\textbf{Notebook scoring rubric.}
    This rubric leverages existing observations from the literature to characterize notebooks along the sensemaking spectrum.
    For example, Tukey's definition of exploratory data analysis motivates our definition of stages 0.1 - 0.5~\cite{Tukey1977}, and we defined stages 0.6 - 1.0 using existing definitions of narrative structure~\cite{rule2018exploration} and types of descriptors~\cite{wang2021makes}.
    Our supplementary material (\url{https://osf.io/9q4wp/?view_only=61e6f58d29194742a0aaed328afdea4d}) includes the full rubric.}
    \label{tab-rubric}
\end{table*}

%% ---------------------------------------------------------------------
%% Study 3
%% ---------------------------------------------------------------------
\section{Measuring Notebook Evolution}
\label{sec:study3}

Rule et al.\ observe that ``the process used to collect, explore, and model data has a significant impact on the sense made.''
In other words, the \emph{process} of authoring a notebook affects the insights derived.
Given that a single snapshot of a notebook represents only one point within this process, it stands to reason that analyzing only one version of a notebook is insufficient to fully comprehend the sensemaking process behind it.
For example, it is impossible to know from a single notebook version whether a user's analysis \emph{shifted} from exploration towards explanation, as hypothesized in prior work, or followed a different path.

However, a more complete view of the user's sensemaking process could be gained by considering how the notebook has changed over time, i.e., across multiple git versions.
%\textit{We hypothesize that computational notebooks will change across versions in a manner that is consistent with our understanding of the sensemaking loop (Pirolli and Card 2005)---each iteration shifting between exploration and explanation.} 
To this end, we analyze how our sensemaking scores from Section~\ref{sec:study2} change across notebook versions by treating them as individual time series.
%Since sensemaking scores capture notebook characteristics \autoref{sec:study2}, changes in notebooks map directly to changes in scores across versions.
%We created a temporal dataset that mapped each observed notebook version on GitHub to a quantitative sensemaking score using our hybrid regression model.
We seek to answer the following research question through this analysis:
\textit{How does the sensemaking score of a notebook change over time, and what factors (if any) may explain any observed changes in the score?}

\subsection{Measuring Sensemaking Scores Across Versions}

To understand how notebooks change over time, we chose to characterize notebooks by their respective and available versions.
This was done in two steps. 

\textit{First}, we used public GitHub commits as a proxy for notebook versions.
We downloaded all available GitHub versions for each notebook.
For each version, we generated the notebook metrics needed to apply the ``hybrid-focused" formula from Section~\ref{sec:study2-hybrid}.
We used these metrics to calculate the sensemaking score of each version, compiling a list of scores for each notebook.
This transformation allowed us to view each notebook as a time-series of sensemaking scores (i.e., a series of notebook scores ordered based on the time of each commit).
For example, we would represent a notebook with 5 versions with a list of 5 numbers, each ranging from 0.1 to 1.0.
Each number indicated the position of each version within the sensemaking spectrum.
We viewed changes in the time-series to indicate the evolution of a notebook across versions. 

\textit{Second}, we normalize the time-series data to enable comparison across notebooks.
The number of versions and thus the length of our representations varied across notebooks, ranging from 4 to 94 versions.
To do this, we generated a simple, best-fit linear regression model for each notebook representing points as a linear relationship between version numbers and sensemaking scores.
A linear model is an appropriate choice because we focus on general shifts across entire notebook histories, which is a noisy time series.

\paragraph{Linear Models}

Many time series exhibit different patterns at different levels of granularity~\cite{heer2006}, where some of the observed variation may be due to noise~\cite{barnett2005}.
The stock market is a classic example.
The gyrations of the stock market vary non-linearly at a granular level, but a linear model can overcome the effects of noise to reveal overall trends of stock market prices over time, e.g., market booms and busts and phenomena such as ``regression to the mean"~\cite{barnett2005}.
A linear model is simple but still appropriate for assessing these kinds of trends in noisy time series~\cite{foster1992}. 

\begin{figure}
    \centering
    \includegraphics[width=0.4\linewidth]{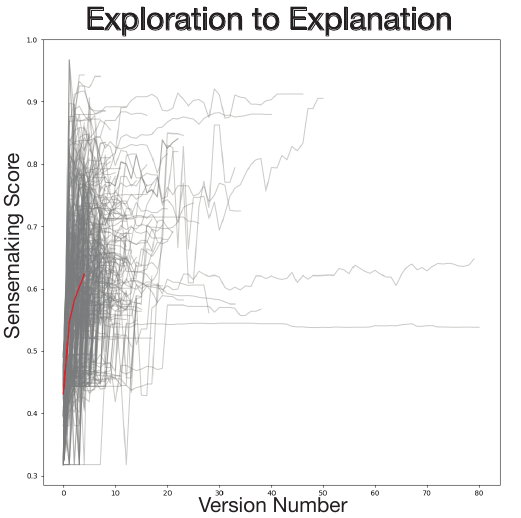}
    \caption{\textbf{Evolution of the Explore-Explain group.}
    This group includes all notebooks that shifted from exploration to explanation, with the average shift plotted in red.} 
    \label{fig:study3-cluster}
\end{figure}

We also attempted to analyze this data using more sophisticated time series analysis methods such as dynamic time warping.
However, we soon realized these methods were unsuccessful due to noise; 
example time series are shown in~\autoref{fig:study3-cluster}.
We observed consistent overall shifts across time series, but no consistent patterns between consecutive pairs of commits.
Hence we adopted a more traditional time series analysis method, i.e., a linear model~\cite{foster1992}.

\subsection{Grouping Time-Series}

Now that we had a means of comparing notebook time-series, we chose to group notebooks by major shifts in sensemaking score as a way to identify common user behaviors. We wanted to asses whether these behaviors matched our current understanding of the sensemaking spectrum.
For example, if users generally follow the pattern hypothesized in prior work~\cite{rule2018exploration,kery2018story, kery2017variolite, dong2021splitting, head2019messes}, then we would expect to see notebooks shifting upward from exploration towards explanation.
However, the notion of a sensemaking loop suggests that users might also do the reverse, corresponding to shifts from explanation toward exploration. 
In the remainder of this section, we describe our process for grouping the time series and qualitatively analyzing each group, and discuss the major shifts that users tended to make along the sensemaking spectrum.

%We group notebook time-series by their first and last scores to identify shifts in sensemaking.
\paragraph{Grouping Methods} We focus our analysis on how notebooks \emph{shift} along the sensemaking spectrum, represented by three variables: initial score (i.e., time-series starting with high/low sensemaking scores), final score (i.e., time-series ending with high/low sensemaking scores), and direction of slope from respective linear regression model (i.e., increasing or decreasing scores). 
%First, we identify whether the scores for the first and last notebook versions were exploratory or explanatory.
Using the rubric established in \autoref{sec:study2}, we labeled scores $\leq{}0.5$ as exploratory and $>0.5$ as explanatory.

We identified four main groups of notebook shifts: exploration to exploration, exploration to explanation, explanation to explanation, and explanation to exploration.
For example, time-series that began and ended with exploratory notebook versions were grouped as 'exploration to exploration'.
%In this way, we can see how well existing hypotheses align with our observations. For example, if sensemaking tends to shift from exploration towards explanation, then we should see a large group of `exploration to explanation' notebooks.

\paragraph{Qualitative Analysis Methods}
Three of the authors qualitatively examined a random sample of 5\% of all notebooks (142 total) and their version histories using the following guidelines:

\begin{enumerate}
    \item We analyzed the first version to form a hypothesis for the notebook author's initial intent in creating the notebook.
    \item We observed changes in the type of text, code, and visualizations across individual version deltas and how these changes contributed to the notebook's narrative.
    \item We paid special attention to changes in the structure of the notebook across versions---e.g., markdown, comments, or visualizations demarcating different analytical steps.
    \item The frequency of commits, the commit window, and the commit messages gave our coders clues into how authors leveraged GitHub to meet their analysis goals.
\end{enumerate}

We derived qualitative codes (words or short phrases) to describe our observations with respect to these guidelines.
We used these codes to identify broader behavioral themes within each of the four sensemaking groups.
Themes focus on structural elements commonly used to track the narrative and flow of sensemaking, including code comments, objectives, sections, templates, and cleaning~\cite{rule2018exploration}.
Details are provided in our supplementary material.

\subsection{Results}

\begin{figure}
    \centering
    \includegraphics[width=\columnwidth]{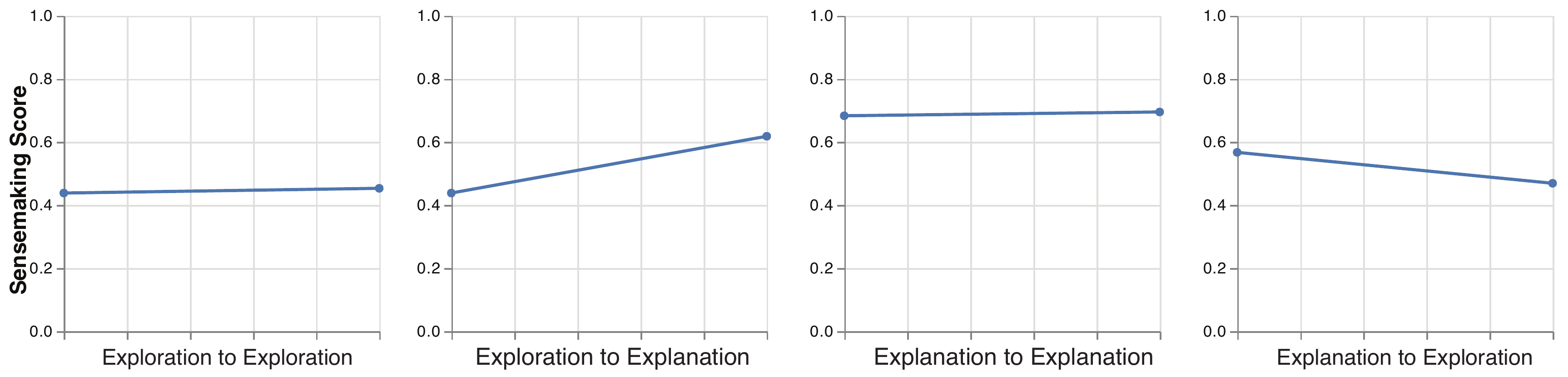}
    \caption{\textbf{Visual summary of notebook time-series.}
    Summary within each notebook group, in order: exploration to exploration, exploration to explanation, explanation to explanation, and exploration to exploration.} 
    \label{fig:study3-flow}
\end{figure}

Here we discuss our observations for each group of sensemaking shifts, summarized in \autoref{tab-group} and \autoref{fig:study3-flow}: exploration to exploration, exploration to explanation, exploration to explanation, and explanation to exploration.

%Groups were constructed based on the first and last scores of the time-series corresponding to notebooks.
%We will define and discuss four groups: exploration to exploration, exploration to explanation, exploration to explanation, and explanation to exploration. A more in-depth qualitative study of the sample helped identify the unique editing behavior of notebook authors within each group. We will discuss our observations within each section.  

\subsubsection{Exploration to Exploration}

22.6\% of our sample contains notebooks that begin as exploratory (0.31-0.49) and remain exploratory after subsequent changes (scores of 0.31-0.49).
%These notebooks demonstrate very narrow and slow-changing shifts within the exploratory part of the sensemaking spectrum. \leilani{what does ``very narrow and slow-changing'' mean? How is speed assessed?}
Notebooks in this group tend to have a relatively flat slope, suggesting ``slow'' progress along the sensemaking spectrum.
Although they remain exploratory, we still observe both positive (towards explanation) and negative (towards exploration) shifts within this group.
% 58.4\% of these notebooks shift with a positive slope, starting with a sensemaking score of 0.42 and ending with a score of 0.45. 38.3\% of these notebooks shift with a negative slope, starting with an average sensemaking score of 0.46 and ending with 0.44. Notebooks that shift neutrally (3.26\%) stay at the sensemaking score of 0.44. 

\paragraph{Edit Behavior}

Authors of these notebooks often depend on \emph{code comments} to organize, annotate and save code~\cite{rule2018exploration, kery2018story, kery2017variolite}.
Code is commented as a way to control the flow of the analysis~\cite{rule2018exploration, kery2018story, kery2017variolite}.
Authors also add text within code comments to label analyses and describe insights. 
These notebooks organize code based on their purpose.
Code in loops and functions are sometimes found in separate cells from code that outputs text, tables, or visualizations.
Code that outputs are generally found in smaller sections to facilitate quick iteration~\cite{kery2018story}. 
The edit behavior within negatively and positively sloping notebooks are the same.
Negatively sloping notebooks often capture the removal of visualizations and positively sloping notebooks their additions. 

\subsubsection{Exploration to Explanation}

15.1\% of our sample contains notebooks that begin as exploratory (scores of 0.31-0.49) and become explanatory (0.50-0.91).
%These notebooks are the fastest-growing notebooks in our sample
These notebooks had a relatively steep slope, which could be interpreted as ``rapid'' shifts in sensemaking.
They tend to start within a narrow range of exploration scores and end within a wider range of explanatory scores.

\paragraph{Edit Behavior}

The first versions of these notebooks typically contain just code or code and visualizations.
In subsequent versions, there are two main methods of iteration authors employ.
The first method consists of adding markdown and code in tandem, such as including annotations and headers into sections as they create and edit code cells.
In the second method, authors focus on code iteration first, and add markdown and headers in their last few commits~\cite{rule2018exploration, kery2018story, dong2021splitting}.
Some authors explicitly label a \emph{cleaning phase} within their GitHub versions where they prep their notebook for communication purposes.
This cleaning phase often involves reordering, splitting, and reformatting code cells as well as adding observations within markdown cells~\cite{dong2021splitting, head2019messes}. 

\subsubsection{Explanation to Explanation}

60.1\% of our sample contains notebooks that begin as explanatory notebooks (scores of 0.50-1.00) and remain explanatory after changes (0.50-1.00).
Although both positive and negative shifts are observed within this group, these notebooks tend to have relatively flat slopes, similar to the `exploration to exploration' group.
This is by far the largest group observed, suggesting that data scientists may prioritize clarity and reproducibility during sensemaking within notebooks, consistent with prior work~\cite{rule2018exploration, pimentel2019large, kery2018story, dong2021splitting, head2019messes}.

\paragraph{Edit Behavior}

Notebooks that became less explanatory (sloped negatively) often began with a \emph{template, to-do list, or a statement of objective} at the top of the page.
In other words, they started as highly explanatory, which we see reflected in these notebooks' first scores, averaging 0.7.
Notebooks that became more explanatory (sloped positively) lacked an explicit statement of objectives.
Objectives, implicit (in positively sloping notebooks) or explicit (in negatively sloping notebooks), seemed to drive the construction of the rest of the notebook.
For example, if a template specified three goals, we observed authors attempt each goal sequentially across versions.
Some authors even added commit messages about the goal being achieved.
When authors implemented each goal, they often added annotations to describe and explain their workflow as it progressed.
For example, if authors added code and visualizations to the end of the notebook, they also added markdown text to describe their process and results.

\subsubsection{Explanation to Exploration}

Perhaps not surprisingly, only 2.09\% of notebooks started explanatory (scores of 0.50-0.74) and became exploratory (0.33-0.49). 
Relative to other groups, these notebooks shift negatively from within a narrow explanatory range to a narrow exploratory range.

\paragraph{Edit Behavior}

These notebooks progress towards exploration through the removal of explanatory elements.
For example, several notebook authors commented code producing visualizations and deleted markdown cells in later versions.
This reduction of visualizations and markdown in favor of code may be indicative of authors preparing for new iterations of sensemaking with existing (and likely duplicate) code as a starting point~\cite{koenzen2020code, kery2017variolite}. 

\subsection{Summary}

%When represented as shifts in sensemaking scores, we observe that notebook edits seem to reflect corresponding shifts within the sensemaking loop. Further, our qualitative observations of notebook edit-behavior appear to be consistent with previous work~\cite{kery2018story, rule2018exploration, dong2021splitting, Head:2019}. 

Our qualitative findings suggest that GitHub commits can capture shifts in notebook editing behaviors over time, which we successfully mapped to corresponding shifts in the authors' sensemaking.
Thus, our results support the idea that one can \emph{automatically} detect a variety of sensemaking activities within computational notebooks.

Although we do see the shift from exploration to explanation emphasized in prior work~\cite{kery2018story, rule2018exploration, dong2021splitting,  head2019messes}, our analysis also reveals a variety of shifts along the entire sensemaking spectrum which were previously unobserved.
For example, data scientists may explain their findings \emph{in tandem} with exploring their data, as seen through our analysis of the `explanation to explanation' group of notebooks.
Furthermore, the `exploration to exploration' group shows that some notebooks have yet to reach the explanatory stage, suggesting that some authors are content to keep certain analyses exploratory.

We also observed shifts \emph{away} from explanation towards exploration.
Though previously unobserved (and in some ways, counter-intuitive), this result is consistent with our understanding of the sensemaking spectrum.
We believe that some of these notebooks demonstrate the beginning of a new sensemaking iteration.
The fact that some of these notebooks start with explicit objectives suggests that authors begin these notebooks with prior knowledge and analysis goals and likely leverage them to streamline their analysis of the data. 
%Our results show that in observing shifts in sensemaking scores, we can \emph{automatically} identify a variety of sensemaking activity within notebooks. In abstract, these are the cumulative shifts across the sensemaking loop.

\begin{table*}[htbp]
    \centering
    \begin{tabular}{cp{1.1cm}p{1.1cm}p{1.1cm}p{1.1cm}}
    \toprule 
    &
    \textbf{Explore-Explore} &
    \textbf{Explain-Explain} &
    \textbf{Explore-Explain} &
    \textbf{Explain-Explore} \\
    \toprule
    \# of notebooks       & 582   & 1549   & 390   & 54     \\
    \% of sample          & 22.6 & 60.1  & 15.1 & 2.00   \\
    avg \# of versions          & 9     & 10     & 10    & 11     \\
    avg first score       & 0.438 & 0.683  & 0.438 & 0.567  \\
    avg last score        & 0.453 & 0.695  & 0.618 & 0.469  \\
    avg slope value       & 0.002 & 0.0015 & 0.021 & -0.012 \\
    \% positively sloping & 58.4 & 54.4  & 96.9 & 5.55 \\
    \% negatively sloping & 38.3 & 45.5  & 3.00  & 94.4 \\
    \% neutral sloping    & 3.20 & .06 & 0     & 0 \\
    \bottomrule
    \end{tabular}
    \caption{\textbf{General statistics of each notebook group.}
    Explore signifies that the score (first-last) in the time-series corresponding to the notebooks are in exploratory side of the sensemaking spectrum.
    Explain signifies that the score is in explanatory side of the spectrum.}
    \label{tab-group}
\end{table*}

%% ---------------------------------------------------------------------
%% Discussion/Implications
%% ---------------------------------------------------------------------
\section{Discussion}
\label{sec:discussion}

We have presented an analysis of 60,000 Jupyter Notebooks and their respective GitHub histories.
With this corpus, we isolate 2,574 notebooks that appear to be data science-oriented, characterize their organization and structure, quantitatively measure various properties to situate them within the overall sensemaking process~\cite{pirolli2005sensemaking}, and observe how sensemaking within these notebooks shifts across GitHub commits.

\subsection{Explaining and Generalizing the Results}

Our results demonstrate that we can apply qualitative observations from the literature (e.g., \cite{rule2018exploration, kery2018story, head2019messes, dong2021splitting, wang2021makes, wang2022documentation}) to \textit{automatically} measure sensemaking within Jupyter Notebooks. 
By measuring how each data science notebook evolves across multiple GitHub commits, we can estimate how and whether subsequent revisions correlate with a notebook becoming more explanatory over time as hypothesized in prior work.
Shifts in scores enabled us to identify the discrete steps of analysts as they iterate within the sensemaking spectrum.
As Pirolli and Card describe, analysts appear to exhibit a cycle of sensemaking activities.

% \deepthi{"how exactly does this work enrich our collective understanding of how people make sense of data?"}
% \leilani{what exactly was corroborated? Do we have a specific example we can point to? Maybe that people often have a goal in mind when they explore, shown with the behavior of the explanation-exploration and explanation-explanation groups.}

% Our findings corroborate results from prior studies on data exploration~\cite{evaBattle, alspaugh2018futzing, liu-alternatives, kandel-enterprise, kandogan-insight, feng-patterns-pace, muller2019data, muller2019human, kery-towards}, showing that some patterns observed in that prior work can also be observed in computational notebooks.
% Thus, we believe our work can be directly applied towards creating more data-driven, interactive, interfaces and tools. 

Our findings also reveal a range of distinct notebook edit behaviors, such as setting analysis goals prior to notebook development. 
These behaviors align with existing observations of data exploration behaviors with visualization tools~\cite{evaBattle}, suggesting that there may be core patterns to sensemaking that seem to transcend particular tools and environments.
%Thus, we believe our work can be directly applied towards creating more data-driven, interactive, interfaces and tools. 
As a result, our work opens the door to gaining a deeper quantitative understanding of the sensemaking loop itself through the lens of data science tools and practices.

\subsection{Implications for Data Science Tool Design}

Our findings show that authors often use structural aspects of the notebook to track and manage the evolution of their analysis (Section~\ref{sec:study3}).
For example, notebook authors will use markdown cells to annotate sections of analysis with their objectives.
We believe that this finding highlights a need for tools that help data scientists manage their goals \emph{while} they perform analysis in a notebook~\cite{head2019messes}. 

% \paragraph{Support Goal Management}
% We observed notebook authors using markdown headers, text, and code comments to impose a goal-oriented document structure before performing data analysis (see \autoref{sec:study3}). 
% Notebook platforms could support this behavior by enabling \emph{linking} between stated objectives and sections of the notebook. A mechanism to create collapsible sections of cell groups would also enable authors to work within the confines of individual goals effortlessly.

\paragraph{Track Multiple Analysis Paths}

% Branches of exploratory analysis and the notebook's analysis workflow were often managed using code comments. 
% In contrast to code comments, we suspect that traditional versioning (like GitHub) does not support the quick and dynamic creation of new exploratory analysis flows~\cite{kery2017variolite, head2019messes, kery-messes}.
% Thus, we recommend implementing a means of encouraging faster and more dynamic versioning on traditional versioning platforms.
% For example, notebook metadata could track alternative analysis paths by explicitly tracking cell dependencies.
% When changes to cell dependencies are versioned on traditional version control schemes, authors could reverse engineer metadata history to find previous analysis paths. This information can be used to review and share results. 
% This method causes minimal disturbance to the analysts' workflow and the extensible notebook environment. 

GitHub versioning of computational notebooks does not help to track what data scientists do.
For example, an analyst may pursue a particular line of inquiry, realize that a few analysis steps were dead ends, and backtrack to an earlier point to continue their analysis---introducing an alternative branch of investigation.
It is hard to represent this non-linear flow with GitHub commits.
We need mechanisms that track the actual non-linear and iterative practices of data scientists~\cite{kery2017variolite, head2019messes, kery-messes}.
We suggest that an extension to current computational notebooks could remedy this problem---an extension that versions and manages cell dependencies.
This enhancement would enable notebook users to better track their sensemaking processes and enable researchers to study sensemaking (and its evolution) in notebook environments.

% Then, we  can talk about how we can combine this with traditional version control schemes to enable systems to reverse engineer a user's previous analysis paths. Using the reverse-engineered data, we can render the analysis history for users, which they can leverage to review and share their work.}

\paragraph{Generate Relevant Recommendations}

Using the techniques we have demonstrated, notebook platforms can automatically calculate the position of a notebook document within the sensemaking spectrum \emph{while it is being edited}. 
Platforms could use this information to support or even enforce best practices.
For example, having detected that the author is in the exploratory phase of analysis, the platform may choose to automatically version the document to comprehensively capture competing branches of exploration.

We believe this information can be particularly pertinent to data science engines that wish to guide analysts with recommendations on analysis tools and techniques.
For example, having detected that an author is performing exploratory analysis, a recommendation engine can cull recommendations from a group of curated \emph{exploratory} notebooks.
Our ideas can direct how recent work, done by Yan et al.~\cite{Yan2020} and Raghunandan et al.~\cite{raghunandan2021}, generate recommendations found in Jupyter Notebooks.
They can use the context of a notebook to provide more targeted data science recommendations to authors. 

\subsection{Limitations and Future Work}

Although our techniques produce a relatively small sample compared to the original corpus, our study is still one of the largest analyses of Jupyter Notebooks from GitHub (e.g., compared to \cite{wang2021makes, dong2021splitting}).
Part of the problem is the inconsistent notebook quality on GitHub~\cite{wang2021makes}.
We combat this challenge by proposing a method to identify data science notebooks suitable for quantitative analysis.
This methodology could easily be extended to collect larger notebook corpora in the future; for example, by curating data science notebooks from all of the millions of notebooks on GitHub.

We approached our dataset with an understanding that many authors selectively report their analysis~\cite{liu2020paths}.
As our findings indicate, many notebooks on GitHub are skewed towards the explanatory side of the spectrum, suggesting that some authors may wait until later in the sensemaking process to share their notebooks.
Coupled with a lack of ground truth for the mental models of the notebook authors, our ability to infer user intent was limited.
We note that this is a fundamental limitation of surveying computational notebooks stored in a public repository such as GitHub, but that the benefits of getting the kind of insight demonstrated here far outweighs this drawback.
We address this limitation in part through a mixed-methods analysis strategy in \autoref{sec:study2} and \autoref{sec:study3}.

Nevertheless, it would be interesting to develop new strategies for collecting richer notebook metadata to fill observed gaps in GitHub histories and to infer user intent from this metadata.
We view our work in this paper as the first of many to explore mixed methods towards understanding sensemaking in computational notebooks.

%% ---------------------------------------------------------------------
%% Acknowledgments
%% ---------------------------------------------------------------------
\section*{Acknowledgments}

%Anonymized for double-blind review.

We gratefully acknowledge the classic 1998 Konami game \textit{Dance Dance Revolution} for inspiring our paper title.
Allow us to close with the following immortal words drawn from this game: ``\textit{You're a rockstar. You're the one they came to see. I'm crying, buckets of tears.}''

%% ---------------------------------------------------------------------
%% References
%% ---------------------------------------------------------------------
\bibliographystyle{ACM-Reference-Format}
\bibliography{note-evolve}

\end{document}